\documentclass{JHEP}
\usepackage{amssymb}


\preprint{{\tt hep-th/0405063}}

\newcommand{\be}{\begin{equation}}
\newcommand{\ee}{\end{equation}}
\newcommand{\bea}{\begin{eqnarray}}
\newcommand{\eea}{\end{eqnarray}}

\title{Chiral effective potential in ${\cal N}=\frac{1}{2}$
non-commutative Wess-Zumino model}

\author{ A.T. Banin ${}^1$, I.L. Buchbinder${}^2$, N.G. Pletnev${}^{1}$\\
~~~~~~~~\\
${}^1$ Institute of Mathematics, \\
Novosibirsk, 630090, Russia,\\
~~~~~~~~\\
${}^2$ Tomsk State
Pedagogical University \\
 Department of Theoretical Physics, \\
Tomsk, 634041, Russia\\
~~~~~~\\
\email{joseph@tspu.edu.ru, pletnev@math.nsc.ru}}

\abstract{ We study a structure of holomorphic quantum
contributions to the effective action for ${\cal N}=\frac{1}{2}$
noncommutative Wess-Zumino model. Using the symbol operator
techniques we present the one-loop chiral effective potential in a
form of integral over proper time of the appropriate heat kernel.
We prove that this kernel can be exactly found. As a result we
obtain the exact integral representation of the one-loop effective
potential. Also we study the expansion of the effective potential
in a series in powers of the chiral superfield $\Phi$ and
derivative $D^2{\Phi}$ and construct a procedure for systematic
calculation of the coefficients in the series. We show that all
terms in the series without derivatives can be summed up in an
explicit form. }

\begin{document}

\section{Introduction}
The deformation of superspace and construction the Moyal superstar
product based on nontrivial (super)Poisson manifolds \cite{A}, as
a first step in constructing supersymmetric string field theory,
has been attracted much attention. It has been shown
\cite{ew} that in the language of the string field theory the Moyal
product is the simplest description of bosonic
string interactions, corresponding to strings joining or splitting. In the
recent literature there are other fundamental studies of
a star product in a certain class of quantum field
theories on noncommutative (NC) Minkowski space-times, which origin
is the Seiberg-Witten limit of open strings in the presence of an
external constant NS-NS B-field \cite{nsew}. To be more precise,
by wrapping the branes with non-zero constant background
field $B_{mn}$ we get the corresponding low energy effective gauge theory
which is deformed to a noncommutative (super)symmetric gauge
theory in such a way that (bosonic) directions become
noncommutative. This result generated a modern activity in
studying quantum field theories in NC space (see \cite{dn},
\cite{ag}, \cite{aref} for reviews). Also we point out that
noncommutative field theories provide a simplified ground for
investigating the nonlocal string theory effects.

Interplay between noncommutative field theories (NCFT) and string
theories has been a rich and fruitful source for better
understanding of both. Part of the interest has been aimed at
getting the new insights into the regularization and
renormalization of quantum field theories in this novel framework
that are neither local nor Lorenz invariant. Another significant
interest is conditioned by the relation between classical and
quantum nonlocality in these theories. In fact, many features of
ordinary (commutative) field theories have found rich analogues in
the NC context. The first revealed unexpected property of NCFT
\cite{mrs} is that even in massive theories there is (UV/IR)
mixing of ultraviolet and infrared divergences. As a consequence,
the Wilsonian approach to field theory seems to break down:
integrating out hight-energy degrees of freedom produces
unexpected low-energy divergences.

For the reasons mentioned above the elementary quanta in NCFT are
no longer point particles; instead of it, the physical excitations
are described by the NC-dipoles --- weakly interacting, one
dimensional extended objects \cite{rey}. In a generic NCFT always
exists a special class of composite operators: open Wilson lines
$W_k(\Phi)$ and their descendants $(\Phi W)_k(\Phi)$. Infrared
dynamics of NC dipoles and hence the open lines is dual to
ultraviolet dynamics of the elementary fields $\Phi$'s.

Furthermore in NC gauge theories (NCGT), which appear naturally in
various decoupling limits of the worldvolume theories of D-branes
in background NS-NS B-field, the gauge invariance becomes subtle.
The translations along NC directions are a subset of gauge
transformations and thus there are no gauge invariant local
operators in the position space. It turns out that NC gauge theories
allow a new type of gauge invariant objects which are localized in
the momentum space (see for reviews \cite{dn, rey, liu}).

The open Wilson line is defined in terms of a path-ordered
$\star$-product, and its expansion in powers of the gauge
potential involves a generalized $\star_n$-product at each $n$-th
order. Generalization of the $\star$-product appears in the
one-loop effective action of NCGT \cite{krsy, z}, in couplings
between the massless closed and open string modes \cite{gar}, and
in Seiberg-Witten map between the ordinary and NC Yang-Mills
fields \cite{wess}. So it has been found a complete agreement of the
results with the Seiberg-Witten limit of the string world-sheet
computation and standard Feynman diagrammatic at low-energy and the
large noncommutative limit \cite{krsy, z}.

More recently, the extension of noncommutativity to odd variables
has been related to the presence of other background fields. In
particular, in \cite{ov} it has been studied the $C$-deformation of
${\cal N}=1$ supersymmetric gauge theories in four dimensions and
computed the coupling to the graviphoton superfield
$F^{\alpha\beta}$ arising from the higher-genus amplitudes in the
(topological) superstring theory. They introduced the
noncommutativity only in the Grassmann odd coordinates, which
breaks spacetime supersymmetry explicitly. New type of a deformation
has been proposed by Seiberg \cite{s} who introduced
noncommutativity both in Grassmann even and odd coordinates but
imposed the commutativity in the chiral coordinates. This
deformation is made by such a way that the anticommuting
coordinates $\theta$ form a Clifford algebra
\begin{equation}\label{deform}
\{\hat{\theta}^{\alpha},\hat{\theta}^{\beta}\}=2 \alpha'^2
F^{\alpha\beta} = C^{\alpha\beta}~.
\end{equation}
The other commutation relations are determined by the
consistency of the algebra. In
particular, the ordinary spacetime coordinates $x^{m}$ can not commute
\begin{equation}
[\hat{x}^{m},\hat{\theta}^{\alpha}]=i
C^{\alpha\beta}\sigma^{m}_{\beta
\dot{\alpha}}\bar{\theta}^{\dot{\alpha}}~,\quad
[\hat{x}^m,\hat{x}^n] = \bar{\theta}\bar{\theta}
C^{mn}~,\quad
\{\bar{\theta}^{\dot{\alpha}},\bar{\theta}^{\dot{\beta}}\}=0~,
\end{equation}
where
$C^{mn}=C^{\alpha\beta}\varepsilon_{\beta\gamma}\sigma^{mn\,\gamma}_{~~\alpha}.$
In contrast to the spacetime coordinates, the chiral coordinates $y^{m}=x^{m}+
i\theta^{\alpha}\sigma^{m}_{\alpha\dot{\alpha}}\bar{\theta}^{\dot{\alpha}}$
can be taken commuting. Note that because the anticommutation
relation of $\bar{\theta}$ remains undeformed,
$\bar{\theta}$ is not the complex conjugate of $\theta$, that is
possible only in the Euclidean space.  The product of functions of
$\theta$ on the chiral superspace is Weyl ordered by using the
star-product, which is the fermionic version of the Moyal product:
\begin{equation}\label{star}
f(\theta)\star g(\theta) = f(\theta)~ {\rm exp}\left(-\frac{1}{2}
~C^{\alpha\beta}
\frac{\overleftarrow{\partial}}{\partial\theta^{\alpha}}
\frac{\overrightarrow{\partial}}{\partial \theta^{\beta}}\right)~
g(\theta)~.
\end{equation}

The supercharges are defined as follows
\begin{equation}
Q_{\alpha} = i\frac{\partial}{\partial \theta^{\alpha}}~, \quad
\bar{Q}_{\dot{\alpha}}= i(\frac{\partial}{\partial
\bar{\theta}^{\dot{\alpha}}}-i\theta^{\alpha}\frac{\partial}{\partial
y^ {\alpha \dot{\alpha}}})~.
\end{equation}
We use the conventions of \cite{ggrs} and therefore we expect
$Q_{\alpha}$ have to be symmetry generators on the ${\cal
N}=\frac{1}{2}$ supersymmetry. The star-product (\ref{star}) is
invariant under the action of supercharges $Q_{\alpha}$. However,
because $\bar{Q}_{\dot{\alpha}}$ depends explicitly on $\theta$,
it is clear that the star-product is not invariant under
$\bar{Q}$. Such a deformation saves the ${\cal N}=\frac{1}{2}$
supersymmetry and has interesting properties in the field theory
viewpoint. Replacing all ordinary products with the above
$\star$-product, one can proceed studying a supersymmetric field
theory in this non(anti)commuting superspace taking into account
that this deformed supersymmetry algebra admits well-defined
representations. Namely, one can define chiral and vector
superfields much similarly to the ordinary ${\cal N }=1$
supersymmetry \cite{s}. Some recent papers deal with various
aspects of field theories defined on such NC superspace \cite{s},
\cite{g}-\cite{lr}. More recently, also the instanton
configurations of the ${\cal N} = \frac{1}{2}$ gauge theory have
been analyzed \cite{ll}. Theories with nilpotent deformation of
${\cal N}=2$ supersymmetry have been constructed in \cite{flm}.

It is very interesting to study how the deformation (\ref{deform})
modifies the quantum dynamics of supersymmetric field theories,
paying particular attention to consequences of nonlocality in the
superspace caused by Eq.(\ref{star}). It was pointed that the
deformation (\ref{star}) induces local operators multiplied by the
non(anti)commutativity parameter $C^{\alpha \beta}$ \cite{s}.
However, even though $C$ breaks Lorentz invariance, the theory
with an arbitrary superpotential $W(\Phi)_{\star}$ is Lorentz
invariant because the deformation depends only on ${\rm det}(C)$.
These induced operators have higher scaling dimension (${\rm dim}
= 6$), and they might cause the theories to be nonrenormalizable.

Though new kinds of (anti)chiral superfields in ${\cal
N}=\frac{1}{2}$ supersymmetric theory violate the holomorphy, the
anti-holomorphy still remains. For deformed WZ-model, this leads
to the non-renormalization theorem of the anti-chiral
superpotential and vanishing of the vacuum energy. Moreover, one
can show that such deformed theories have finite number of
divergent structures in their effective actions and hence, they
are in fact renormalizable. Besides, unlike the ${\cal N}=1$
supersymmetric models, in ${\cal N}=\frac{1}{2}$ models containing
the chiral and antichiral superfields exist the loop corrections
to the chiral effective action even in a massive case (see for
comparison a situation in standard ${\cal N}=1$ SUSY models
\cite{CHP} and in WZ model with noncommutative spacetime
\cite{CHP1}.

One more important observation was that the deformed WZ-model
after adding the new $F$ and $F^2$ terms (which are not written in
the star deformation) to the original lagrangian becomes
renormalizable up to two loops \cite{g}. The renormalizability was
then extended to all orders in the perturbation theory for the
deformed Wess-Zumino model in \cite{bf}, \cite{ro}, and for
deformed gauge theories with(out) matter in \cite{lr}. In
particular, these works show that, thought these terms carry
scaling dimensions large than four, the deformation-induced
operators do not lead to power divergences in loop diagrams
because of absence the Hermitian conjugate operators to them and
they are radiatively corrected at least logarithmically. Also it
was shown that all divergent terms have one power insertion of
${\rm det}(C)$ and counterterms $F, F^2, F^3$ suffice to
renormalized the theory. Using nonstandard scaling dimension
assignment, in Ref. \cite{br} has been explain intuitively
renormalizability of the NCFT. With insertions of all possible
undeformed and deformed terms having the dimensions less then or
equal to four in the lagrangian, the theory under consideration is
multiplicatively renormalizable.

Other remarkable quantum properties of supersymmetric theories
such as stability of the vacuum energy and the existence of an
antichiral ring, remain unchanged despite of the fact that NC induced
a soft-breaking of supersymmetry. The vacuum energy remains zero
to all orders in perturbation theory. Structure of the effective
action in the ${\cal N}=\frac{1}{2}$ WZ model has been studied in
\cite{f} where an analysis for summing the one-loop contributions
to the effective potential has been presented for the case of small
odd and even external momenta.  Unfortunately, it is unclear how
to interpret the physical sense of such approximations.

In this work we develop a general approach to constructing the
one-loop effective potential in ${\cal N}=\frac{1}{2}$ WZ model.
The approach is based on use of the symbol operator techniques and
heat kernel method and allows to carry out a straightforward
calculation of one-loop finite quantum corrections. As a result we
find an exact form of one-loop effective potential for the
considered model in terms of a proper-time integral. Also we
construct a new scheme for approximate evaluation of the effective
potential and give a complete solution of the problem settled up
in \cite{f}.

The paper is organized as follows. Section 2 is devoted to a
formulation of the model. In Section 3 we describe a general
procedure of calculations. Subsection 3.1 is devoted to a brief
discussion of the symbol operator techniques and its application
to finding the one-loop effective action for the theories in
${\cal N}=1$ superspace. In Subsect 3.2 we calculate the exact
heat kernel and find the one-loop effective potential for the
theory under consideration. The effective potential provides the
complete solution of the problem formulated in \cite{f}.
Subsection 3.2 is devoted to constructing a scheme for approximate
calculation of the effective potential in a form of an expansion
in the field $\Phi$ and derivatives $D^2{\Phi}$. In Section 4 we
present a procedure for an explicit calculation of the terms in
the above expansion. Section 4.1 is devoted to finding the
divergences in the framework of our general scheme. Explicit
calculations of two first finite terms in the expansion of the
effective potential are presented in the Subsection 4.2.1 and
Subsection 4.2.2. It is interesting to point out that these terms
as well as the others can be expressed in terms of hypergeometric
functions. In Subsection 4.3 we construct an approximation form of
the effective potential by summing up all terms including all
powers of superfield $\Phi$ but do not containing the derivatives
$D^2{\Phi}$. Summary is devoted to discussion of the results
obtained.

\section{The model}
On the ${\cal N}=\frac{1}{2}$ noncommutative superspace the WZ
model is given by the standard classical action where the point
products of superfields are replaced with the star product
(\ref{star}):
\begin{equation}\label{cact}
S = \int d^8z\, \bar{\Phi}\star \Phi + \int d^6z
\,(\frac{m}{2}\Phi \star \Phi +\frac{g}{3!}\Phi \star \Phi \star
\Phi) + \int d^6\,\bar{z}(\frac{\bar{m}}{2}
\bar{\Phi}\star\bar{\Phi} +
\frac{\bar{g}}{3!}\bar{\Phi}\star\bar{\Phi}\star\bar{\Phi})~.
\end{equation}
Chiral superfields are defined by the relation
$\bar{D}_{\dot{\alpha}}\Phi(y,\theta,\bar{\theta}) =0$, which
means $\Phi(y,\theta)= A(y)+\theta\kappa(y) +\theta^2 F(y)$.
Anti-chiral superfields are defined by the relation $D_{\alpha}
\bar{\Phi}(y,\theta,\bar{\theta})=0$, which means
$\bar{\Phi}(\bar{y},\bar{\theta})= \bar{A}(\bar{y})
+\bar{\theta}\bar{\kappa}(\bar{y}) +\bar{\theta}^2
\bar{F}(\bar{y})$. As it has been demonstrated in Ref. \cite{s},
the $\star$-product of the chiral superfields is again a chiral
superfield; likewise, the $\star$-product of the antichiral
superfields is again an antichiral superfield.

The model is formulated in Euclidean space where the fields $\Phi,
\bar\Phi$ are considered as independent. Using the property $\int
\Phi \star \Phi =\int\Phi\cdot\Phi$, performing the expansion of
the star-product (\ref{star}) and turning down total superspace
derivatives, the cubic interaction terms reduce to the usual WZ
interactions complemented by the terms violating ${\cal N}=1$
supersymmetry to ${\cal N}=\frac{1}{2}$ supersymmetry.
\begin{equation}\label{action}
\begin{array}{c}
S=\int d^8z \,\bar{\Phi} \Phi + \int d^6z \,(\frac{m}{2}\Phi\Phi +
\frac{g}{3!}\Phi\Phi\Phi) +\int d^6\bar{z}\,(
\frac{\bar{m}}{2}\bar{\Phi}\bar{\Phi}
+\frac{\bar{g}}{3!}\bar{\Phi}\bar{\Phi}\bar{\Phi})+\\ \\
+ \int d^6z\, ( \frac{h}{3!}\Phi (Q^2\Phi)^2
+\frac{1}{2\lambda}\Phi (Q^2\Phi))~,
\end{array}
\end{equation}
where $h=-\frac{g}{4}|\det C|$. Last term containing the coupling
$\lambda$ is added to provide a multiplicative renormalization of
the model (see e.g \cite{bf}). As a result we see that
the action (\ref{cact}) is rewritten in terms of
standard ${\cal N}=1$ superspace, i.e. without star-product.
Hence, one can consider the deformed WZ model as
ordinary WZ model, where superfield multiplication is
standard, with a new addition to the $F$-term.

Thus we treat the theory as some special model formulated in
terms of ${\cal N}=1$ superspace and this circumstance allows us to use
all the standard tools and techniques of superspace quantum field
theory.

It is important to note that superpotential $W$ is connected with
the K\"ahler potential. This is because $\bar{\theta}^2
=\delta(\bar{\theta})$ is a chiral superfield and the K\"ahler
term $\int d^4\theta \bar{\theta}^2K(\Phi,\bar{\Phi})$ can be
converted to the chiral superpotential $\int d^2\theta K(\Phi,
\bar{A})$. Thus one can expect that the non-renormalization
theorem for the superpotential is no longer true and actually we
will see that there exist the quantum corrections to the
superpotential of the deformed WZ model\footnote{Structure of the
K\"ahlerian effective potential in the standard WZ model is
discussed in \cite{bky}.}.

The new (non)renormalization theorems have been proven
\cite{bf}-\cite{t}: the $F$-term is radiatively corrected and
becomes indistinguishable from the $D$-term, while the
$\bar{F}$-term is not renormalized. Since the supersymmetric vacua
are critical points of the antiholomorphic superpotential, the
vacuum stays stable. New divergences arise from sectors containing
non-planar diagrams (\cite{g}, \cite{bf}, \cite{ro}, \cite{b}).
This is because the planar diagrams do not depend on the deformed
parameter except those for the star product between the external
legs.

The superfields $\Phi,\bar{\Phi}$ satisfy the classical equation
of motion
\begin{equation}\label{eqofmot}
\begin{array}{c}
D^2\Phi + \bar{m}\bar{\Phi}
+\frac{1}{2}\bar{g}\bar{\Phi}^2=0~,\\
\bar{D}^2\bar{\Phi}+m\Phi
+\frac{1}{2}g\Phi^2+\frac{h}{2}(Q^2\Phi)^2+\frac{1}{\lambda}(Q^2\Phi)=0~,
\end{array}
\end{equation}
which are the algebraic equations for the auxiliary fields $F$ and
$\bar{F}$
\begin{equation}\label{mot} -F= \bar{G}\equiv\bar{m}
\bar{A}+\frac{1}{2}\bar{g}\bar{A}^2~,\quad -\bar{F}=
G+\frac{h}{2}F^2+\frac{1}{\lambda}F~.
\end{equation}

In component form, the effect of the star deformation leads to an
additional $F^3$ term in the action in comparison with the
standard WZ model. It means that the only scalar potential is
affected by the deformation. The potential expressed in components
fields is
\begin{equation}\label{bpot} V= \bar{G}(G
+\frac{1}{\lambda}\bar{G}+\frac{h}{6}\bar{G}^2)~, \end{equation} where
$G=m A+ \frac{g}{2}A^2$.
By solving the equations
$\frac{\partial V}{\partial A}=\frac{\partial V}{\partial
\bar{A}}=0, V=0$, ones find a set of the ${\cal N}=\frac{1}{2}$
supersymmetric vacua with vanishing vacuum energy.

It has been shown (see e.g. Refs. \cite{s}, \cite{g}, \cite{bf},
\cite{f}, \cite{b}, \cite{t}) that at one-loop a divergent term
proportional to $F^2$ appears, which is not present in the
classical action. To make the theory multiplicatively
renormalizable we consider a modified action with the addition of
$F^2$ term with coupling $\lambda$ in (\ref{action}). The
radiatively generated $F^2$-term is interesting because this the
term gives rise to mass splitting between the boson and fermion
component fields, while keeping the vacuum energy to zero. The
appearance of $F^2$ divergence might lead to the conclusion that
the star-product becomes deformed at the quantum level. Indeed, in
Ref. \cite{f} it has been given a general argument allowing to
prove that suitable resummation of such and other terms in the
effective potential (in some limit of large noncommutativity
parameter and small external momenta) can be rewritten in terms of
open Wilson lines.

\section{General scheme of calculating the one-loop effective potential }
In this section we describe a calculation of the one-particle
irreducible (IPR) effective potential for the model
(\ref{action}). The one-loop correction to the effective action is
formally written in the form
\begin{equation}
\Gamma_{(1)} = {i\over 2}\ln {\rm Det} (\hat{H})~,
\end{equation}
where $\hat{H}$ is the differential operator acting on superfields
being the second variational derivatives over quantum
(super)fields of the classical action. In order to find this
operator in the framework of the loop expansion one have to split
all fields of the model on quantum and background parts. We use
the standard quantum-background splitting
$$
\Phi \rightarrow \Phi + \varphi, \quad \bar{\Phi} \rightarrow
\bar{\Phi} + \bar{\varphi}~,
$$
where $\Phi$ and $\varphi$ stand for background and quantum fields
respectively.   The rules for calculating variational derivatives
$\varphi, \bar{\varphi}$ are analogous to the ones given in Refs.
\cite{ggrs}, \cite{bk}:
$$
\frac{\delta\varphi(z)}{\delta\varphi(z')}=\bar{D}^2\delta(z-z')~,
\quad
\frac{\delta\bar{\varphi}(z)}{\delta\bar{\varphi}(z')}={D}^2\delta(z-z')~.
$$
We will keep both $m, \bar{m}$ nonzero so that the effective
action is well defined in the infrared. The quadratic over quantum
(super)fields part of the classical action is written in the form
\begin{equation}
S_{(2)} = \frac{1}{2}\int d^8z\; (\bar{\varphi}, \varphi)\hat{H}
\pmatrix{\varphi \cr\bar{\varphi}}~,
\end{equation}
where we denote
\begin{equation}\label{svari}
\hat{H}= \pmatrix{ D^2\bar{D}^2&(\bar{m}+\bar{g}\bar{\Phi})D^2\cr
(m + \Lambda)\bar{D}^2&\bar{D}^2D^2 }~,\quad  \Lambda =  g\Phi +
h(Q^2\Phi)Q^2 + \frac{1}{\lambda}Q^2~.
\end{equation}
Further we use the convenient denotations
\begin{equation}\label{repl}
m+g\Phi=\mu,\quad \bar{m}+\bar{g}\bar{\Phi}=\bar{\mu}~
\end{equation}
and consider the special background $\bar\Phi = Const$, $\bar\mu =
Const$ which is sufficient for calculation of the chiral effective
potential (see a discussion in Refs. \cite{s}-\cite{lr}).

Let's present the operator $\hat{H}$ in action $S_{(2)}$ as a sum
of two operators
\begin{equation}\label{hath}
\hat{H}=\hat{H_0}+\hat{H_1}~,
\end{equation}
where the operators are defined as follows

\begin{equation}
\hat{H_0}=\pmatrix{ D^2\bar{D}^2 &\bar{\mu}D^2\cr m\bar{D}^2
&\bar{D}^2D^2
 },\;
 \hat{H_0}^{-1}=\pmatrix{ \frac{D^2\bar{D}^2}{\Box}\frac{1}{\Box
-\bar{\mu}m
}&\frac{-\bar{\mu}}{\Box-m\bar{\mu}}\frac{D^2}{\Box}\cr
\frac{-m}{\Box-\bar{\mu}m}\frac{\bar{D}^2}{\Box}&\frac{\bar{D}^2D^2}{\Box}\frac{1}{\Box
-m\bar{\mu}}},
 \hat{H_{1}} =\pmatrix{0,& 0\cr \Lambda \bar{D}^2& 0}.
\end{equation}
From (\ref{hath}) using the relation $\ln (\hat{H_0}+\hat{H_1}) =
\ln (\hat{H_0}) + \ln (1 + \hat{H_0}^{-1}\hat{H_1})$ we extract
the contribution of $H_{0}$. The structure of the expression ${\rm
Tr}\ln (\hat{H_0})$ corresponds to an unbroken ${\cal N}=1$
supersymmetry. This expression does not depend on the superfield
$\Phi$ and can be omitted. As a result ones get the expression for
the chiral effective potential
\begin{equation}\label{mmu}
\Gamma_{(1)}=\frac{i}{2}{\rm Tr}\ln\left(1+\pmatrix{
\frac{-\bar{\mu}}{\Box-m\bar{\mu}}\frac{D^2\bar{D}^2}{\Box}\Lambda&
0\cr \bar{D}^2\frac{1}{\Box-m\bar{\mu}}\Lambda& 0 }\right) =
\frac{i}{2} {\rm Tr}\ln \left( 1 -
\frac{\bar\mu}{\Box-m\bar{\mu}}\frac{D^2\bar{D}^2}{\Box}\Lambda\right)~.
\end{equation}
All operators $\Box, D, \bar{D}$ in this expression act trough,
i.e. not only on $\Lambda$. Now we expand the logarithm and
obtain a series in powers of $\frac{D^2\bar{D}^2}{\Box}\Lambda$,
which can be transformed as follows
$$
\frac{D^2\bar{D}^2}{\Box}\Lambda\frac{D^2\bar{D}^2}{\Box}\Lambda\cdots
=
\frac{D^2}{\Box}\Lambda\frac{\bar{D}^2D^2\bar{D}^2}{\Box}\Lambda\cdots
= \frac{D^2}{\Box}\Lambda\bar{D}^2\Lambda\cdots
=\frac{D^2\bar{D}^2}{\Box}\Lambda^2\cdots~,
$$
here we take into account that $\Lambda$ is a chiral field (i.e.
$[\bar{D}, \Lambda]=0$) and the identity
$\frac{\bar{D}^2D^2\bar{D}^2}{\Box} = \bar{D}^2$. Therefore the
identity
$$
\left(\frac{D^2\bar{D}^2}{\Box}\Lambda\right)^n\cdots =
\frac{D^2\bar{D}^2}{\Box}\Lambda^n \cdots
$$
takes place and one can write (\ref{mmu}) as follows
\begin{equation}\label{mmu1}
\Gamma_{(1)}=\frac{i}{2}{\rm
Tr}\left(\frac{D^2\bar{D}^2}{\Box}\ln(1-\frac{\bar{\mu}}{\Box-m\bar{\mu}}\Lambda)\right)~.
\end{equation}

In order to simplify (\ref{mmu1}) we rewrite it in the form
\begin{equation}\label{mmu2}
\begin{array}{rl}
\Gamma_{(1)}&=\displaystyle\frac{i}{2}{\rm
Tr}\frac{D^2\bar{D}^2}{\Box}\ln(\frac{\Box-m\bar{\mu}-\bar{\mu}g\Phi-\bar{\mu}\tilde{\Lambda}}{\Box-m\bar{\mu}})\\
&=\displaystyle\frac{i}{2}{\rm
Tr}\frac{D^2\bar{D}^2}{\Box}\left(\ln(\Box-\mu\bar{\mu}-\bar{\mu}\tilde{\Lambda})
-\ln({\Box-m\bar{\mu}})\right)~,
\end{array}
\end{equation}
here
$$
\tilde{\Lambda}= h(Q^2\Phi)Q^2+\frac{1}{\lambda}Q^2= MQ^2~.
$$

The quantity $\ln({\Box-m\bar{\mu}})$ in the second line
(\ref{mmu2}) can be dropped out since it doesn't depend on the
superfield $\Phi$. Thus, the one-loop contribution to the
effective potential (\ref{mmu}) is finally presented in the
following form
\begin{equation}\label{g0}
    \Gamma_{(1)} =\frac{i}{2}{\rm
Tr}\left(\frac{D^2\bar{D}^2}{\Box}\ln(\Box-\mu\bar{\mu}-\bar{\mu}\tilde{\Lambda})\right)
~.
\end{equation}
We pay attention on appearance of the chiral projector in this
relation showing that the effective action is given by an integral
over a chiral subspace. Further calculations will be fulfilled
using the symbol-operator techniques \cite{bbp}.

\subsection{Symbol-operator techniques and heat kernel representation}
Here we shortly describe the basic notions of the symbol-operator
techniques which is used for calculations of the one-loop
effective action. The detail description of this techniques was
given in papers \cite{bbp}. We just remind the main stages.

The main idea is based on the supersymmetric generalization of
the well known trace formula for the operator $\hat{A} = a(\hat{q},
\hat{p})$
\begin{equation}\label{str} {\rm Tr} (\hat{A})= \int
d\mu(\gamma) A(\gamma)~,
\end{equation}
where $\hat{q}, \hat{p}$ are the operators of coordinate and momentum,
$\gamma = (q, p)$ are the coordinates on the phase-space,
$d\mu(\gamma)$ is a measure on the phase-space, $A(\gamma)$ is a symbol
of the operator $\hat{A}$ and integration goes over the full
phase-space. The symbol of the operator $\hat{A}$ is function
on phase space and defined\footnote{The symbol we use here is so called
$pq$-symbol. One can show that ${\rm Tr} (\hat{A})$ does not depend on
the choice of symbol.} as $A(\gamma) = \frac{\langle
p|\hat{A}|q\rangle}{\langle p|q\rangle}$.  Practical calculation of
symbols is based on a relation like $A(\gamma) =
a(q+i{\hbar}\frac{d}{dp}, p) \times 1$ where right hand side is
constructed from the operator $a(\hat{q}, \hat{p})$ replacing the
operator $\hat{q}$ by $q + i{\hbar}\frac{d}{dp}$ and the operator
$\hat{p}$ by $p$ (see e.g.\cite{bbp} and the references therein).

We apply the symbol operator techniques to calculation of traces
for the operators depending on ${\cal N}=1$ superspace coordinates
$z^{M} =(x^{m},\theta^{\alpha}, \bar{\theta}^{\dot{\alpha}})$ and
corresponding derivatives. The phase superspace is parameterized
by $z^{M},p_{M}$ where $p_{M} = (p_{m}, \psi_{\alpha},
\bar{\psi}_{\dot{\alpha}})$. In the case under consideration we
have to consider the operators $\hat{A}= a(z,
\frac{\partial}{\partial z})$.  According to the symbol-operator
techniques, the calculation of the trace for such operators
contains the following steps (see the details in \cite{bbp}):

{\bf 1}. We introduce the quantity $\langle p|z\rangle$
and define the symbols $(z, p)$ of the basic operators
$(z, \frac{\partial}{\partial z})$ by general relation $(z, p)\langle
p|z\rangle = \langle p|(\hat{z}, \hat{p})|z\rangle$.

{\bf 2}. To calculate a symbol of the operator function $a(z,
\frac{\partial}{\partial z})$ of basic operators ones replace all
basic operators $z, \frac{\partial}{\partial z}$ by the quantities
$z^{\hbar}, p$ respectively. The quantities $z^{\hbar}$ are
constructed by the rule $z^{\hbar} = U^{-1}zU$. The appropriate
operator $U$ for ${\cal N}=1$ superspace field theories is found
in \cite{bbp} and will be written down bellow.

{\bf 3}. Symbol of the operator under consideration is
defined as $A(\gamma) = a(z^{\hbar}, p)\times 1$.

{\bf 4}. Trace of the operator $\hat{A}$ under consideration is given
by relation (\ref{str}).

Let us employ the above prescription for calculating the effective
action (\ref{g0}). In this case the phase superspace coordinates
are $\gamma = (p, \psi,\bar\psi; x, \theta, \bar\theta)$, the
measure in the trace definition (\ref{str}) is $d\mu(\gamma)=
d^8z\, \frac{d^4p}{(2\pi)^4} d^2\psi d^2\bar{\psi}$ and the
quantity $\langle p, \psi,\bar\psi|x, \theta, \bar\theta\rangle$
is
\begin{equation}
\langle p, \psi,\bar\psi|x, \theta,
\bar\theta\rangle = {\rm e}^{ip \cdot
x+\theta^{\alpha}\cdot\psi_{\alpha}+\bar{\theta}^{\dot{\alpha}}\cdot
\bar{\psi}_{\dot{\alpha}}}~.
\end{equation}
It allows to obtain
the symbols of the supercharge $Q_{\alpha}$ and spinor derivatives
$D_{\alpha}, \bar{D_{\dot{\alpha}}}$ in
the form
\begin{equation}\label{symb}
\begin{array}{rcl} {\rm\bf
Operator}& &{\rm\bf Symbol}\\ Q_{\alpha} =
i\frac{\partial}{\partial\theta^{\alpha}}=i\partial_{\alpha}
&\rightarrow &Q_{\alpha}(\gamma) =
i\psi_{\alpha}~,\\
D_{\alpha}=\partial_{\alpha}
+\bar{\theta}^{\dot\alpha}i\partial_{\alpha \dot\alpha}
&\rightarrow &D_{\alpha}(\gamma) = \psi_{\alpha} -
\bar{\theta}^{\dot\alpha}p_{\alpha
\dot\alpha}~,\\
\bar{D}_{\dot\alpha}=\bar{\partial}_{\dot\alpha}&\rightarrow &
\bar{D}_{\dot\alpha}(\gamma) = \bar{\psi}_{\dot\alpha}~.
\end{array}
\end{equation}

In order to calculate the symbol of the operator function $a(z,
\partial_{m}, D, \bar D, Q)$ we have to replace within the
operator the $z, \frac{\partial}{\partial z}$ by the corresponding
quantities  $z^{\hbar}, p$ constructed with the help of a
$U$-operator which, for the case under consideration, has the form
(see the details in \cite{bbp})
\begin{equation}
U = {\rm e}^{-\bar{\partial}\bar{D}} {\rm e}^{\partial p
\cdot\bar{\theta}}{\rm e}^{-\partial D}{\rm
e}^{-i\partial_{p}\partial_{x}},
\end{equation}
where $\partial^{\alpha}= \frac{\partial}{\partial \psi_{\alpha}},
\bar{\partial}^{\dot\alpha}= \frac{\partial}{\partial
\bar{\psi}_{\dot\alpha}}$ and the derivatives $D, \bar{D},
\partial_{z}$ act on the left while the operators
$\frac{\partial}{\partial {\psi}},
\frac{\partial}{\partial{\bar{\psi}}}, \partial_{p}$ act on the
right. 
As a result ones get the operator
acting in phase superspace\footnote{Index ${\hbar}$ marks the
quantities transformed with the help of $U$-operator.}

\begin{equation}
A^{\hbar} = a(z^{\hbar}, p^{\hbar}, D^{\hbar}, \bar{D^{\hbar}},
Q^{\hbar})~.
\end{equation}
Simple calculations lead to
\begin{equation}\label{sp_der}
\bar{D}^{\hbar}_{\dot\alpha}=\bar{\psi}_{\dot\alpha}~, \quad
D^{\hbar}_{\alpha} = \psi_{\alpha} -
\bar{\partial}^{\dot\alpha}p_{\alpha\dot\alpha}~, \quad
Q^{\hbar}_{\alpha}= i(\psi_{\alpha} +
p_{\alpha\dot\alpha}\bar{\theta}^{\dot\alpha})~, \quad
p^{\hbar}_{\alpha\dot\alpha} = p_{\alpha\dot\alpha}~.
\end{equation}
These operators obey the initial algebra but act on functions of
phase superspace coordinates. For the background field one can get
the representation
\begin{equation}\label{sp_field}
\Phi^{\hbar}(y, \theta) = \Phi(y+i\partial_p,
\theta+\partial_{\psi})= \Phi +
\partial^{\alpha}(D_{\alpha}\Phi) -\partial^2(D^2\Phi)+ {\cal
O}(\partial_y)~,
\end{equation}
here the derivatives $\partial= \frac{\partial}{\partial\psi}$ act
through on the right.
 The terms including the $y$-derivatives of $\Phi$ can be
omitted because we use the approximation of background fields
slowly varying in space-time.

Now we consider the effective action (\ref{g0}). It is written as
a trace of the operator depending on $z, D. \bar D, Q$ and hence,
we can apply to its calculation the symbol operator techniques we
shortly described above.  After replacement in (\ref{g0}) the
derivatives and fields by corresponding quantities (\ref{sp_der})
and (\ref{sp_field}) we obtain the representation for the one-loop
effective action in the form:
\begin{equation}\label{gam5}
\Gamma_{(1)} = \frac{i}{2}\int d^8z \,\int d^2\psi d^2\bar{\psi}\,
\frac{d^4 p}{(2\pi)^4} \left(\frac{\psi^2 \bar{\psi}^2}{-p^2}\right)
\ln(-p^2 - \bar{\mu} \cdot (m + g\Phi^{\hbar}(y, \theta)+ M
(Q^{\hbar})^ 2))\times 1~,
\end{equation}
\begin{equation}\label{bigm}
 M= h(Q^2 \Phi) + \frac{1}{\lambda}~.
\end{equation}
We replace in (\ref{g0}) the operator $\frac{D^2\bar{D}^2}{\Box}$
with (\ref{sp_der}) and take into account that the non-zero
integral over $\bar{\psi}$ must contain the factor
${\bar{\psi}}^2$. Since such a factor is completely formed in
$\bar{D}_{\hbar}^2$ and saturates the integral over $\bar{\psi}$,
we can drop $\bar{\partial}^{\dot\alpha}p_{\alpha\dot\alpha}$ in
$D_{\hbar}^2$.

 Last step of calculations is to act on unit in
(\ref{gam5}). It is equivalent to moving all derivatives
$\partial_{\psi},
\partial_{\bar\psi}$ in (\ref{gam5}) on the right and drop them.
The properties of the superspace integration and differentiation
\begin{equation} \int
d^2\psi \,A \equiv\partial^2 A, \quad \partial^2\cdot \psi^2 = -1
\end{equation}
can be used in (\ref{gam5}) for calculation.

Useful elements of calculations of the effective action under
consideration is $\zeta$-function representation of Eq.
(\ref{gam5}). To find it we use proper-time integral
representation of the logarithm. It is convenient also to
introduce a dimension parameter $L^2$ which serves for
compensation of the dimensional quantity in the proper-time
exponent\footnote{The proper time $T$ became dimensionless.}. With
the mentioned notations the $\zeta$-function representation is
\begin{equation}
\Gamma_{(1)}=\left.\left(-\frac{d}{ds}\right)\right|_{s=0}\frac{i}{2}
\int d^8z \,\int\frac{d^4p}{(2\pi)^4}\int d^2\psi d^2\bar{\psi}\,
\left(\frac{\psi^2 \bar{\psi}^2}{-p^2}\right)
\int^{\infty}_{0}\frac{d T}{\Gamma(s)}T^{s-1}(L^2)^s e^{-T(p^2
+\bar{\mu}\mu + \bar{\mu} \tilde\Lambda)}=\label{ga1}
\end{equation}
$$=\left.\left(-\frac{d}{ds}\right)\right|_{s=0}\int^{\infty}_{0}\frac{d T}{\Gamma(s)}\,T^{s-1}\int d^6z \,
\bar{D}^2 K(\frac{T}{L^2})~,
$$
where the heat kernel in the above equation is defined as
\begin{equation}\label{hker}
K(T)= \frac{i}{2}\int \frac{d^4p}{(2\pi)^4}\,\frac{1}{-p^2}
\left(\frac{L^2}{\bar{\mu}}\right)^s \bar{\mu}  \; {\rm e}^{-T(p^2
+ m)}{\rm e}^{-T (M Q^2_{\hbar} + g\Phi_{\hbar}(y, \theta))}\times
1 ~.
\end{equation}
It should be noted that our aim is study only the chiral effective
potential and, therefore, we rewrite the integral over
grassmannian variables in (\ref{ga1}) via a chiral integral.

In (\ref{ga1}), for simplicity, we changed the denotation for
variables:
\begin{equation}\label{pchange}
p^{2}/\bar{\mu}\rightarrow p^{2}~, \quad T\bar{\mu}\rightarrow T~.
\end{equation}
In further calculation we have to keep in mind these replacement.

As it has been shown by direct calculation in Refs. \cite{rey},
\cite{krsy}, \cite{z}, \cite{ro}, \cite{lr} and presented as a
theorem in \cite{b} the chiral superpotential in ${\cal
N}=\frac{1}{2}$ model is proportional to $\bar{\theta}^2$. The
presence of the term $\sim\Phi (Q^2\Phi)^2$ in the classical
action (\ref{action}) leads to possible action of $Q^2$ inside the
loop integrals, and this results effectively in
$\bar{\theta}^2$-dependence.

The general structure of the effective action in ${\cal
N}=\frac{1}{2}$ supersymmetry models is given as \cite{b}
\begin{equation}
  \Gamma[\Phi, \bar{\Phi}]=\sum_{n}\int \prod_{j=1}^{n}d^4x_{j}\,
  \int d^2\theta d^2\bar{\theta}\,
  G(x_{1},\ldots,x_{n};\bar{\theta}^2)F_{1}(x_1,\theta,\bar\theta)\cdots
  F_{n}(x_n,\theta,\bar\theta)~,
\end{equation}
where $G(x_{1},\ldots,x_{n};\bar{\theta}^2)$ are
translation-invariant functions of coordinates $x$ and possible
insertion $\bar{\theta}^2$, while $F(x,\theta,\bar\theta)$ are
local operators of $\Phi, \bar{\Phi}$, their covariant
derivatives, and the expressions $Q\Phi$. The insertion of
$\bar{\theta}^2$ from $G(x_{1},\ldots,x_{n};\bar{\theta}^2)$
absorbs the $\int d^2\bar{\theta}$ integral. Because of this, the
$D$-terms with pure chiral fields and holomorphic $F$-terms
coincide between each other and, therefore, both $D$-terms and
$F$-terms are unified in ${\cal N}=\frac{1}{2}$ supersymmetry
\cite{b}.

As a result, the integral over full superspace (\ref{ga1}) is
transformed to the integral over chiral subspace $\int d^8z
\bar{\theta}^2\ldots =\int d^6z \bar{D}^2\bar{\theta}^2 \ldots$.
Therefore quantity $\bar\mu$ will contribute to the effective
action (\ref{hker}) only one of it's component $\bar\mu=\bar{m}
+\bar{g}\bar{A}$ Using equation of motion it means $\bar{\mu}^2=
\bar{m}^2 - 2\bar{g} F$. Thus, the expression for chiral potential
(\ref{ga1}) does not contain any antichiral superfields.

\subsection{Exact calculation of the heat kernel}
In the present section we present a method allowing to find an
exact expression for the heat kernel. In general, exact
calculation of the heat kernel is impossible. The model under
consideration is quite remarkable since it provides the exact
evaluation of the heat kernel. The reason is the fact that for
this model the heat kernel calculation is reduced to finding an
evolution operator for a harmonical oscillator with the
grassmannian coordinate $\psi$ and momentum
$\partial/\partial\psi$.

In order to calculate (\ref{hker}), according to the symbol-operator
techniques, we have to disentangle derivatives
 in the exponent of the heat kernel. To do that
we transfer all derivatives $\partial_{\psi}$ on the right and act
on unit. It means that after such a transformation all terms with
derivatives must be omitted and a final contribution is resulted
only from recommutations of the differential operators to the last
right position. The rest part is the symbol of the heat kernel.
Finally, the trace of hear kernel is given by integration of the
corresponding symbols over the phase space (see e.g. \cite{bbp}).

Let's denote the right exponent in (\ref{hker}) as
\begin{equation}\label{ht}
  h(T)= {\rm e}^{\displaystyle -T (M Q^2_{\hbar} + g\Phi_{\hbar}(y, \theta))}\times 1~.
\end{equation}
Using (\ref{sp_field}) one can rewrite it in the form
\begin{equation}\label{hamil}
h (T) = {\rm e}^{\displaystyle TM \tilde{Q}_{\hbar}^2 -Tg\Phi -Tg
\partial^{\alpha}(D_\alpha \Phi) +Tg
\partial^2(D^2 \Phi)}
\end{equation}
where $\partial^\alpha
=\frac{\partial}{\partial \psi_\alpha}$, $\tilde{Q}_\alpha^{\hbar} =(\psi +
\sqrt{\bar{\mu}}p\bar{\theta})_\alpha$ and $M$ was
defined in (\ref{bigm}).

The expression in the exponent (\ref{hamil}) can be simplified by
introducing a new denotation $\tilde{\partial}_\alpha =
\partial_\alpha -\frac{D_\alpha \Phi}{D^2\Phi}$, then we can extract from
(\ref{hamil}) $\psi$- and $\partial_{\psi}$-independent part
\begin{equation}\label{kheat}
h (T) =  e^{-Tg\Phi -Tg \frac{(D\Phi)^2}{D^2\Phi}} \cdot
k(T)\times 1~, \quad k(T) = {\rm e}^{TM\tilde{Q}^2_{\hbar} +Tg (D^2 \Phi)
\tilde{\partial}^2}~.
\end{equation}

Straightforward calculations of the commutation relations between
operators in operator-dependent part $k(T)$ lead to the following
algebra
\begin{equation}
\{\tilde\partial^\alpha ,\tilde{Q}_\beta^{\hbar}\}=\delta^\alpha_\beta
\quad [\tilde\partial^2, \tilde{Q}^2_{\hbar}] = (\tilde{Q}_\alpha^{\hbar}
\tilde\partial^\alpha -1)~,
\end{equation}
i.e. $\tilde{Q}_\beta^{\hbar}$ and $\tilde\partial^\alpha$ can be
considered as the Grassmann coordinates and momenta.

We will show that the function $k(T)\times 1$ can be calculated
exactly. It means that the whole heat kernel in (\ref{hker}) is
exactly found. We pay attention that the operator $k (T) = {\rm
e}^{TM\tilde{Q}^2_{\hbar} + Tg (D^2 \Phi)\tilde{\partial}^2}$ can
be evidently treated as an evolution operator for a quantum system
with Hamiltonian ${\cal H} = - ( M\tilde{Q}^2_{\hbar} +g (D^2
\Phi)\tilde{\partial}^2)$. This Hamiltonian is a quadratic form in
grassmannian coordinates $\psi$  and corresponding momenta
$\frac{\partial}{\partial\psi}$ with some coefficients. Such a
quantum system is called the grassmannian oscillator (see e.g.
\cite{FV}). As a result, the problem under consideration is
reduced to a problem of finding the trace for the evolution
operator of the grassmannian oscillator which can be solved
exactly on the base of its algebraic properties.

Let us introduce denotations for the operators
\begin{equation}\label{gendif}
 e_1 = \tilde{Q}^2_{\hbar}, \quad e_2 =\tilde\partial^2, \quad e_3 =
 \tilde{Q}_\alpha^{\hbar}
\tilde\partial^\alpha -1~.
\end{equation}
It is easy to see that these operators satisfy the commutation
relations for the generators of $su(2)$ algebra
\begin{equation}
[e_1,e_2]=e_3, \quad [e_3,e_1]=2e_1, \quad [e_3,e_2]= -2e_2~.
\end{equation}
Hence,
the exponent in $k(T)$ (\ref{kheat}) is nothing but a group
element of $SU(2)$
\begin{equation}
k(T) = {\rm e}^{TMe_1 + TgFe_2}~,
\end{equation}
where
\begin{equation}\label{F}
F=D^2\Phi, M= h(Q^2 \Phi) + \frac{1}{\lambda}~.
\end{equation}
This observation allows to simplify all further consideration.

Since our goal is to find a symbol of the heat kernel, we should
move all derivatives in the exponent (\ref{kheat}) to right hand side
and act on unit what is equivalent to drop them. The
generators containing the derivatives in the group element $k(T)$ are
$e_{2}, e_{3}$.  It is most convenient to rewrite the group element
$k(T)$ in the Gaussian form
\begin{equation}\label{nf}
k(T) = {\rm e}^{TMe_1 + TFe_2} = {\rm e}^{A(T)e_1} {\rm
e}^{B(T)e_3} {\rm e}^{C(T)e_2}~.
\end{equation}
To find functions $A, B, C$ we calculate $\frac{d k(T)}{d T}\cdot k^{-1}(T)$
for first $k={\rm e}^{TMe_1 + TFe_2}$ and for second $k={\rm e}^{A(T)e_1} {\rm
e}^{B(T)e_3} {\rm e}^{C(T)e_2}$ representations for $k$ from the expression (\ref{nf}). Because
the results must be equal, it leads to constrains on the coefficients
\begin{equation}\label{neq}
M e_1 + Fe_2 = e_1(\dot{A} - 2A\dot{B} -
A^2 \dot{C}e^{-2B}) + e_2\dot{C}e^{-2B} +e_3 (\dot{B} +A
\dot{C}e^{-2B})~.
\end{equation}
Comparison of the coefficients at each $e_{i}$ leads to a set of
first order differential equations for $A, B, C$. In particular,
one of these equations looks like $\dot{A} +A^2 F =M$ and has the
solution
\begin{equation}\label{A}
A
=\sqrt{\frac{M}{F}} \tanh(T\sqrt{M F})~.
\end{equation}
The other equations have the solutions
\begin{equation}\label{B}
B= -\ln \cosh (T\sqrt{F M}),
\quad C= \sqrt{\frac{F}{M}} \tanh(T\sqrt{FM})~.
\end{equation}
Here the $F$ and $M$ are given by (\ref{F}).
Using
definition of the generators (\ref{gendif}) and dropping the
derivatives acting on unit in (\ref{nf}) one obtains
\begin{equation}
\begin{array}{rl}
{\rm e}^{A(T)e_1} &= {\rm e}^{A Q^2}~,\\
{\rm e}^{B(T)e_3} &= {\rm e}^{-B(T)} \exp ((1-e^{B(T)}) Q_\alpha
\frac{gD^\alpha \Phi}{F})~,\\
{\rm e}^{C(T)e_2}\times 1 &= {\rm e}^{C(T)\tilde{\partial}^2} =
\exp (C(T) g^2 \frac{D^\alpha\Phi D_\alpha \Phi}{2 F^2} )~,
\end{array}
\end{equation}
where $Q_\alpha = \sqrt{\bar{\mu}}p_{\alpha
\dot\alpha}\bar{\theta}^{\dot\alpha}~, Q^2= -\bar{\mu}p^2
\bar{\theta}^2$ due to (\ref{sp_der}, \ref{pchange}) and only
terms without $\psi, \bar\psi$ are kept because the definition (\ref{ga1})
already contains the factor $\psi^2\bar{\psi}^2$.

As a result one obtains for the symbol (\ref{ht}) the following
expression
\begin{equation}\label{kt}
k(T)\times 1= Q^2\left(A(T) + (1-e^B)^2
\frac{g^2}{F^2}(D\Phi)^2\right) \exp \left(-B(T) +
C(T)\frac{g^2}{F^2}(D\Phi)^2\right)~,
\end{equation}
where $(D\Phi)^2 = \frac{1}{2}D^\alpha\Phi D_\alpha\Phi$
and the $A(T), B(T), C(T)$ are given by (\ref{A}), (\ref{B})

Now using  (\ref{kheat}, \ref{kt})
in (\ref{ga1}, \ref{hker}) we
obtain the exact expression for the one-loop chiral potential
\begin{equation}\label{hexact}
   \Gamma_{(1)} = -\displaystyle\left.{d\over ds}\right|_{s=0} \int
d^6z\int_{0}^{\infty} {dT\over \Gamma (s)}T^{s-1} {1\over 2(4\pi
T)^2}\left(L^2\over \bar{\mu}\right)^{s}\bar{\mu}^2
{\rm e}^{-T(m+g\Phi)}\cdot \tilde{k}(T)~,
\end{equation}
where
\begin{equation}
\begin{array}{l}
\tilde{k}(T) =
\displaystyle\left(\sqrt{M\over F}\tanh(T\sqrt{FM}) +
\left(1-{1\over \cosh(T\sqrt{FM})}\right)^2{g^2\over
F^2}(D\Phi)^2\right)\times\\
\times \cosh (T\sqrt{FM})\exp \left(Tg^2 {(D\Phi)^2\over
F}\left({\tanh(T\sqrt{FM})\over T\sqrt{FM}}-1\right)\right)
\end{array}\label{kfin}
\end{equation}
can be further simplified. Taking into account the grassmannian
property $(D\Phi)^3=0$ and expanding the exponent in (\ref{kfin})
one can find
\begin{equation}
\tilde{k}(T) = \sinh (T\sqrt{FM})\left(\sqrt{\frac{M}{F}}  +
T\sqrt{FM} \left(\frac{\tanh
(T\sqrt{FM}/2)}{T\sqrt{FM}/2}-1\right) \frac{g^2}{F^2}
(D\Phi)^2\right)~.
\end{equation}
Now we take into account the property $\int D^{\alpha}(\Phi
D_\alpha \Phi e^{-Tg\Phi})=0$, which can be used in
(\ref{hexact}). We get $(D\Phi)^2 =-\frac{\Phi F}{1-Tg\Phi}$. As a
result one obtains the expression
\begin{equation}\label{tildek}
\tilde{k}(T) =
 \sqrt{\frac{M}{F}}
\sinh (T\sqrt{FM}) \left(1
-g\frac{Tg\Phi}{1-Tg\Phi}\left(\frac{\tanh
(T\sqrt{FM}/2)}{T\sqrt{FM}/2} -1\right)\right)~. \label{tk}
\end{equation}

The expressions (\ref{hexact}, \ref{tildek}) determine the final
exact solution for the one-loop chiral effective potential in
${\cal N}=\frac{1}{2}$ WZ model. The various approximate results
can be obtained using the various expansions of (\ref{tildek}).
Also we point out that the integral (\ref{hexact}) is divergent at
the low limit. To get a finite effective potential we should, as
usual, to subtract in the integrand of (\ref{tildek}) a first term
in expansion of the integrand in $T$. We will specially discuss a
structure of divergences in the theory under consideration in
Subsection 4.1.

\subsection{Expansion of the heat kernel}
The exact result for the one-loop chiral effective potential is
presented by the expressions (\ref{hexact}, \ref{tildek}) in
the form of an integral over proper time $T$ which can not be written in
an explicit form in terms of elementary or known special functions.
To obtain the various approximate results we have to construct the
expansions of the heat kernel and calculate the integral over
proper time in an explicit form. In this section we formulate some
independent procedure for the heat kernel expansion which doesn't
use (\ref{hexact}) at all and based on the Fourier transforms in the
grassmannian variables.  It allows us to develop an approximate
scheme for calculating the chiral effective potential in a form of a power
expansion of spinor derivatives of $\Phi$. Of course,
such an expansion can be constructed, in principle, on the base of
exact result (\ref{hexact}, \ref{tildek}). However, technically it
is much more easy to begin with initial relation (\ref{hker}).

Let's present the exponent (\ref{ht}) as a sum
\begin{equation}\label{h0v}
h(T)=e^{T(H_0 + V)}\times 1 =\sum_{n=0}^{\infty}h_n(T)\times 1~,
\quad h_0={\rm e}^{T H_0}~, \quad H_{0} =-MQ^2_{\hbar}~, \quad V =
-g\Phi_{\hbar}(y, \theta )~,
\end{equation}
where the general term of the sum is given by the $T$-ordered
iterated integral
\begin{equation}\label{gterm}
h_n(T)\times 1=\int_0^T dt_n \,\int_0^{t_n}
dt_{n-1}\ldots\int_0^{t_2} dt_1\, \, {\rm e}^{(T-t_n)H_0} V {\rm
e}^{(t_n-t_{n-1})H_0} V\ldots V{\rm e}^{(t_2-t_1)H_0} V {\rm
e}^{t_1H_0}~,
\end{equation}
(see details in Ref. \cite{v}). This integral for every fixed $n$
can be calculated. Firstly, we make a replacement of the variables
\begin{equation}
\left\{
\begin{array}{rcl}
s_1 &=& t_2 -t_1~,\nonumber\\
&\cdots~,& \nonumber\\
s_{n-1}&=&t_n-t_{n-1}~,\nonumber\\
t&=& T-t_n~, \nonumber
\end{array}\right.
\end{equation}
using the rules $(t_1,...,t_n) \rightarrow (s_1,...,s_{n-1}, t)$.
This replacement does not change integration because
$\frac{\partial(s_1,...,s_n,t)}{\partial(t_1,...,t_n)}=1$ and
$$
\sum_{i=1}^{n-1} s_i \leq T-t \leq T, \quad 0 \leq s_i, \quad i=1,...,n-1.
$$
Therefore one can write the integral in (\ref{gterm}) as follows
$$
\int_0^T dt_n \int_0^{t_n}dt_{n-1}\ldots\int_0^{t_2}dt_1 = \int
ds_1\ldots ds_{n-1}\int_0^{T-\sum_{i=1}^{n-1}s_i} dt;\quad 0 \leq
s_i, \quad \sum_{i=1}^{n-1}s_i \leq T
$$
Let's introduce a
redundant variable $s_n = T - \sum_{i=1}^{n-1} s_i $, then the
general term (\ref{gterm}) can be written in a symmetrical form
\begin{equation}
h_n(T)\times 1=  \int d^n s\, \delta(T- \sum_{i=1}^{n}s_i)
\int_0^{s_n} ds_{n+1}\, {\rm Tr}\left({\rm e}^{tH_0} V {\rm
e}^{s_1 H_0} ... V {\rm e}^{s_{n-1}H_0} V {\rm e}^{s_nH_0} {\rm
e}^{-t H_0}\right)~.
\end{equation}
Because
$$
\int_0^{s_n} ds_{n+1} = s_n = \frac{1}{n} \sum_{i=1} ^n s_i
=\frac{1}{n} T~,
$$
one can also write
\begin{equation}\label{hsym}
h_n(T)\times 1 = \frac{T}{n} \int d^n s\, \delta(T- \sum_{i=1}^n
s_i) {\rm Tr} \left(V e^{s_1H_0} ... V e^{s_n H_0} \right)~.
\end{equation}

For further calculation we use another replacement of the
variables $s_i \rightarrow \alpha_i, \alpha_i =\frac{s_i}{T}$ in
(\ref{h0v}) along with the Fourier transform for the grassmannian
variables
\begin{equation}\label{v}
V(y,\theta) = -g \Phi_{\hbar} (y, \theta ) = \int d^2 \pi \,(-g
\Phi(y, \pi))\, e^{(\theta + \partial) \pi}~,\quad \partial = \frac{\partial}{\partial\psi}~,
\end{equation}
where the Fourier transform of the background field is
\begin{equation}\label{fourier}
\Phi (y, \pi) = -\int d^2\theta \,{\rm e}^{-\theta \pi} \Phi (y,
\theta)= F(y) + \pi^{\alpha}\kappa_{\alpha}(y) - \pi^2 A(y)~.
\end{equation}
This transformation allows to rewrite the expression (\ref{hsym}) in the form
\begin{equation}\label{hn}
\begin{array}{rcl}
h_n (T)\times 1&=& \int \prod_{i=1}^{n}\, d^2 \pi_i\, [-g
\Phi(\pi_i)]
\frac{T^n}{n} \times \\
&\times& \int_{\alpha_i \geq 0}d^n \alpha\, \delta(1- \sum^n
\alpha_i)\, {\rm e}^{\theta (\sum \pi_i)} E_{(n)}(\alpha_{1},\ldots \alpha_{n})
\end{array}
\end{equation}
where we introduced the new denotation
\begin{equation}\label{tail}
E_{(n)}(\alpha_{1},\ldots \alpha_{n}) = {\rm e}^{\partial \pi_1}
{\rm e}^{T \alpha_1 H_0} {\rm e}^{\partial \pi_2}{\rm e}^{T
\alpha_2 H_0}\times ... \times{\rm e}^{\partial \pi_n} {\rm e}^{T
\alpha_n H_0}\times 1~.
\end{equation}

The last expression can be simplified. Because
$TH_{0}=-TMQ^{2}_{\hbar}$ it is useful temporarily introduce a new
variable $u =-TM$. Using the relation
\begin{equation}
{\rm e}^{\partial \pi} {\rm e}^{\alpha u Q^2_{\hbar}} {\rm
e}^{-\partial \pi}{\rm e}^{\partial \pi} = {\rm e}^{\alpha u
(Q^2_{\hbar} - i\pi Q _{\hbar}- \pi^2)} {\rm e}^{\partial \pi}~,
\end{equation}
where $\pi^2 =\frac{1}{2} \pi^{\alpha} \pi_{\alpha}$ along with
the property  $e^{\partial \pi} \times 1 =1$, we transform the
exponent sequence in (\ref{tail}) to
\begin{equation}
{\rm e}^{\alpha_1 u (Q^2 - i\pi_1 Q - \pi_1^2)} {\rm e}^{\alpha_2
u (Q^2_{\hbar} - i(\pi_1 +\pi_2) Q_{\hbar} - (\pi_1
+\pi_2)^2)}\times \cdots \times{\rm e}^{\alpha_n u (Q^2_{\hbar} -
i(\pi_1 + ... + \pi_n) Q_{\hbar} - (\pi_1 + ... + \pi_n)^2)}~.
\end{equation}
Collecting terms with the same powers $Q_{\hbar}$ one can write
\begin{eqnarray}
E_{(n)}(\alpha_{1},\ldots \alpha_{n})=\exp\left(Q^2_{\hbar} u
(\alpha_1+ ...+\alpha_n) -[\alpha_1 \pi_1 +
\alpha_2(\pi_1 + \pi_2)+ ...+\right.\nonumber\\
\left.+\alpha_n(\pi_1 +...+\pi_n) ]i u Q_{\hbar} - u[\alpha_1
\pi_1^2 + \alpha_2(\pi_1+\pi_2)^2 +...+
\alpha_n(\pi_1+...+\pi_n)^2]\right)\times 1~.\label{exp_tail}
\end{eqnarray}

For the integration over grassmannian variables $\psi$ in
(\ref{ga1}) we must keep in $E_{(n)}$ only terms which don't
contain $\psi$ at all. That means that we can drop all $\psi$ in
further calculation. Because we made changes of variables
(\ref{pchange}) the expression for supercharge now is $Q_{\hbar} =
i\sqrt{\bar{\mu}}\hat{p}\cdot\bar{\theta}$. Using relation
$e^{u(Q^2_{\hbar} - i\Gamma^{\alpha}Q_{\alpha}^{\hbar})}=
e^{u\tilde{Q}^2_{\hbar}} e^{u \Gamma^2}$, new denotations
$\tilde{Q}= Q_{\hbar} - i \Gamma$, $Q^2_{\hbar} =
-\bar{\mu}p^2\bar{\theta}^2$ and properties
\begin{equation}
\bar{D}^2 \tilde{Q}^2| = \bar{\mu} p^2~,\quad \int\frac{d^4
p}{(2\pi)^4} e^{-T p^2} = \frac{i}{(4\pi T)^2}~,
\end{equation}
we transfer the consequence (\ref{exp_tail}) to a final form
\begin{eqnarray}
 &E_{(n)}(\alpha_{1},\ldots \alpha_{n})=\exp\left(TM[\alpha_1\pi_1^2 + \alpha_2(\pi_1+\pi_2)^2  \cdots
+\alpha_n(\pi_1+\ldots+\pi_n)^2]+\right.&\label{tail1} \\
&\left.-TM [\alpha_1\pi_1
+\alpha_2(\pi_1+\pi_2)+\ldots+\alpha_n(\pi_1+...+\pi_n)]^2\right)~,&
\nonumber
\end{eqnarray}
where $u=-TM$. This leads to the following representation for the
one-loop chiral effective potential
\begin{eqnarray}
\Gamma_{(1)}&=\left.\left(-\frac{d}{ds}\right)\right|&\int d^4x\int d^2
\theta\,\int_0^{\infty}\frac{dT}{\Gamma(s)}\frac{T^{s-1}}{2(4\pi
T)^2}(-T M)(\frac{L^2}{\bar{\mu}})^s \bar{\mu}^2 {\rm e}^{-T m
}\times\nonumber\\
& & \sum_{n=0}^{\infty} \int \prod_{i=1}^{n} d^2 \pi_i
\,[-g\Phi(y, \pi_i)]
\frac{T^n}{n}\,{\rm e}^{\theta(\sum_{i=1}^{n}\pi_{i} )} \int_{\alpha_i\geq 0} d^n \alpha\times \nonumber\\
& &\delta(1-\sum^n \alpha_i) E_{(n)}(\alpha_{1},\ldots
\alpha_{n})~.\label{intexp}
\end{eqnarray}
This is the basic relation allowing to develop a scheme for
approximate calculations of the chiral effective potential.

Thus, we obtained the one-loop chiral effective potential in two
forms, in form of exact integral representation (\ref{hexact}) and
in form of expansion (\ref{intexp}). Integral representation
(\ref{hexact}) has independent significance and is one of our main
results here. The expansion (\ref{intexp}) will be used further for
obtaining the various approximate results for the chiral effective
potential.

\section{Evaluations of the chiral effective potential }\label{eval}
Now we evaluate the chiral effective potential for the
model (\ref{action}) using the representation
(\ref{intexp}) and the quantities $E_{(n)}$ which appeared
in the heat kernel representation (\ref{hn}). It is clear that in
general the chiral effective potential should have
the divergent and the finite parts
\begin{equation}
  \Gamma_{(1)} = \Gamma_{(1)}^{div} + \Gamma_{(1)}^{fin} = \int d^6z\,
(W_{\rm div}(\Phi, D^2\Phi) + W_{\rm fin}(\Phi, D^2\Phi))~,
\end{equation}
where $W_{\rm div}$ is a divergent part which is
discussed in the section \ref{divergent}, the finite part $W_{\rm
fin}$ is studied in the sections \ref{structure}, \ref{constant}.

Because $\pi$ are the grassmannian momenta $\pi^3=0$, the
quantities $E_{(n)}$ are the finite polynomials in $\pi_i$ for
each fixed $n$ in (\ref{intexp}). Below we will demonstrate the
calculation of several $E_{(n)}$ in an explicit form. It is worth
pointing out that expression (\ref{intexp}) has a structure of the
well-known analytical representation for the one-loop Feynman
integrals and obviously it could be found summing up the
contributions of all one-loop diagrams with arbitrary number of
external legs.

Let us present (\ref{intexp}) in the following form
\begin{equation}\label{expan}
\Gamma_{(1)} = \sum_{n=0}^{\infty}
R_{n}~,
\end{equation}
where
\begin{eqnarray}\label{rn}
R_{(n)}&=\left.\left(-\frac{d}{ds}\right)\right|&\int d^4x\int d^2
\theta\,\int_0^{\infty}\frac{dT}{\Gamma(s)}\frac{T^{s-1}}{2(4\pi
T)^2}(-T M)(\frac{L^2}{\bar{\mu}})^s \bar{\mu}^2 {\rm e}^{-T m
}\times\nonumber\\
& & \int \prod_{i=1}^{n} d^2 \pi_i
\,[-g\Phi(y, \pi_i)]
\frac{T^n}{n}\,{\rm e}^{\theta(\sum_{i=1}^{n}\pi_{i} )} \int_{\alpha_i\geq 0} d^n \alpha\times \nonumber\\
& &\delta(1-\sum^n \alpha_i) E_{(n)}(\alpha_{1},\ldots
\alpha_{n})~.
\end{eqnarray}
Further it will be shown that relation (\ref{expan}) leads to the
effective potential in the form of expansion in power in
$D^2{\Phi}$ with coefficients dependent on $\Phi$
\begin{equation}
\Gamma_{(1)} = \int d^4x d^2\theta\, W_{\rm eff}(\Phi,
D^2\Phi)~, \quad W_{\rm eff}(\Phi, D^2\Phi) =
\sum_{k=0}^{\infty}W_{k}(\Phi)(D^2\Phi)^k~,
\end{equation}
where $W_{k}(\Phi)$ is an infinite series over powers of $\Phi$
which goes from all $R_{l}$ with $l>k$.

In is worth paying attention again on the expansion (\ref{expan})
which is nothing but an ordinary diagram expansion: each $R_{n}$
corresponds to a contribution from diagram having $n$-legs $\Phi$
on the background of constant $D^2\Phi$. It means that the
corresponding contributions contain the mentioned constant
background dependence. The operator techniques we use here allows
to obtain this expansion by simpler way. A sense of the expansion
(\ref{expan}) can be traced from analysis of Eq. (\ref{intexp}).
According to Eq. (\ref{gterm}) the quantity $E_{(n)}$ defines the
heat kernel expansion. The sum over $n$ in (\ref{intexp}) appeared
from the heat kernel expansion (\ref{h0v}, \ref{hn}). From Eq.
(\ref{gterm}) it is follows that the contribution from $h_{n}$ for
any fixed $n$ goes from $n$ insertion of the background field $V$
(or $\Phi$) (\ref{v}). It means that a term $R_{n}$ in the
expansion (\ref{expan}) could correspond to the contribution from
the diagram having $n$ external legs. Nevertheless, according to
(\ref{repl}, \ref{gam5}) the external field $\Phi$ is also
presented in $\mu$ and in $M$, which are contained in the
propagator. The quantity $M$ consists of two terms: first term
$h(Q^2 \Phi)$ will insert additional background field, while the
second term $1/\lambda$ will not. Finally we can conclude that the
term $R_{n}$ in the expansion (\ref{expan}) consists from the
contributions of several one-loop diagrams: namely from diagrams
having $n, n+1, \ldots, 2n$ external fields $\Phi$.

As one can see from (\ref{tail1})
for practical purposes it is useful to introduce the new variables
(linear combinations of the external momenta)
\begin{equation}\label{l}
\begin{array}{rcl}
l_1&=&\pi_1~, \\
l_2&=&\pi_1+\pi_2~,\\
& &\cdots~,\\
l_n&=&\pi_1 +\ldots+\pi_n~,\\
\end{array}
l_{ij}= l_i -l_j~.
\end{equation}

The inverse transformation is
\begin{equation}\label{pi}
\begin{array}{rcl}
\pi_1&=&l_1~,\\
\pi_2&=&l_2 - l_1 = l_{21}~,\\
\pi_3&=&l_3 - l_2 = l_{32}~,\\
& &\cdots~,\\
\pi_n&=&l_n - l_{n-1} = l_{n\, n-1}~,
\end{array}
\end{equation}
and, therefore, the above transformation leads in (\ref{intexp})
to $\Phi(\pi_{k})\rightarrow \Phi(l_{k\,k-1})$. Using property
$\sum_{i} \alpha_i=1$, in terms of variables (\ref{l}), the
exponent (\ref{tail1}) can be written as
\begin{equation}\label{expon}
E_{(n)}(\alpha_{1},\ldots \alpha_{n}) = \exp
(TM(\underbrace{\alpha_1\alpha_2 l_{12}^2 + \alpha_1\alpha_3
l_{13}^2 +\dots+ \alpha_1\alpha_n l_{1n}^2 + \ldots +
\alpha_{n-1}\alpha_n l_{n-1\,n}^2}_{n(n-1)/2}))~.
\end{equation}
Because $\pi_{i}$ are grassmannian momenta, $(l_{i_{1}i_{2}}^2)^k
=0$ for any $k>1$, the expansion of the exponent for fixed $n$
will lead to a finite polynomial in $\pi_{i}$. The expansion of
the above exponential argument gives $\frac{n(n-1)}{2}$ terms with
different coefficients $\alpha_{i_{1}} \alpha_{i_{2}}$
\begin{equation}\label{sume}
E_{(n)}=1 + \sum_{k=0}^n \frac{1}{k!} \sum_{j}
(k;j_1,...,j_{\frac{n(n-1)}{2}})\prod^{\frac{n(n-1)}{2}}_{i=1}(TM\tilde{\alpha}_{i}
l_{i}^2)^{j_i}~,
\end{equation}
where $\sum_{j}$ goes over all possible combinations $j_{i}=0,\,1$
such that $\sum_{i} j_i=k$ and $i=1,...,\frac{n(n-1)}{2}$;
$(k;j_1,...j_m)$ is the number of ways to put $k=j_1+\ldots+j_m$
different things into $m$ different boxes and $j_{i}=0,\,1$;
$l_{i}=l_{i_{1}i_{2}}$ is the variables (\ref{pi});
$\tilde{\alpha}_{i}=\alpha_{i_{1}}\alpha_{i_{2}}$ with the same
induces as the corresponding $l_{i_{1}i_{2}}$.

For example
\begin{equation}\label{sampl}
\begin{array}{lcl}
& &E_{(2)}=\exp( TM \alpha_1 \alpha_2 l_{12}^2)= 1 + \frac{1}{2}
TM \alpha_1 \alpha_2 \pi_1 \pi_2~,\\
 & & E_{(3)}=\exp
(TM(\alpha_1 \alpha_2 l_{12}^2+\alpha_1 \alpha_3
l_{13}^2 +\alpha_2 \alpha_3 l_{23}^2))= \\
& &1 + TM(\alpha_1 \alpha_2
l_{12}^2+\alpha_1 \alpha_3 l_{13}^2 +\alpha_2 \alpha_3 l_{23}^2)+\\
& & (T M)^2(\alpha_1^2 \alpha_2 \alpha_3 l_{12}^2l_{13}^2 +
\alpha_1 \alpha_2^2 \alpha_3l_{12}^2l_{23}^2 + \alpha_1 \alpha_2
\alpha_3^2 l_{13}^2l_{23}^2)+ \\
& &(TM)^3 \alpha_1^2 \alpha_2^2 \alpha_3^2
l_{12}^2l_{13}^2l_{23}^2~,
\end{array}
\end{equation}
where $l_{12}^2=\pi_2^2=-\frac{1}{2} \pi_2 (\pi_1 + \pi_3),$
$l_{13}^2 = \pi_1^2 =-\frac{1}{2}\pi_1 (\pi_2 + \pi_3),$
$l_{23}^2=\pi_3^2=-\frac{1}{2} \pi_3 (\pi_1 + \pi_2)$.

One can note that the expansions (\ref{sume}) correspond to the
derivative expansion in the coordinate space. All terms in
(\ref{sume}), excluding the first one, contain the momenta $\pi$,
which after the inverse Fourier transform are similar to the one
given in Eq. (\ref{v}) will correspond to a grassmannian
derivation, i.e. $\pi_i\Phi_i\rightarrow D\Phi$. The first term in
(\ref{sume}) doesn't contain $\pi$ and therefore leads to the
contribution without derivatives.

After expansion of the $E_{(n)}$ in a series, the finding of
$R_{n}$ contribution in (\ref{expan}) consists in calculation of
the proper-time and $\alpha$- integrals. The integrals over alphas
can be calculated using expression
\begin{equation}\label{aint}
\int \prod_{i}^{n} d\alpha_{i}\; \delta(1-\sum_{i}^n
\alpha_{i})\,\alpha_1^{j_1}\alpha_2^{j_2}\ldots\alpha_n^{j_n}=
\frac{\Gamma(j_1+1)\Gamma(j_2+1)\ldots\Gamma(j_n+1)}{\Gamma(j_1+j_2+\cdots+j_n+n)}~.
\end{equation}

It can be also useful to fulfil the Fourier transform inverse to
(\ref{fourier}), which is equivalent to the replacement
$\pi_i\Phi_i\rightarrow D\Phi$. It is easy to see that the
structure of a general element of the expansion contains $\Phi^m
(D^2\Phi)^n$. Such a structure belongs to the ring structure which
corresponds to the ring structure discussed in \cite{ss}.

In the following sections we will show that for the model under
consideration any term $R_n$ can be expressed via
hypergeometric functions of $n$ variables. Due to grassmannian
properties of the variables the expansions of these hypergeometric
functions are finite polynomials.

\subsection{Divergent part of effective potential}\label{divergent}
In this section we find the divergent contributions to the chiral
effective potential  using two approaches. The first approach
based on direct calculation of expression (\ref{svari}) while the
second approach uses Eq. (\ref{intexp}). Of course, both
approaches give the coincident results.

There are at least two possibilities for calculation of the
divergencies. The first one is based on direct expansion of the
logarithm in (\ref{svari}) into power series of the matrix, i.e.
$\ln(1+X)= X-X^2/2+ X^3/3 -X^4/4+ \ldots $, and calculating the
traces for corresponding matrix powers. For the simplicity we
rewrite matrix $X=H_0^{-1}H_{1}$ from (\ref{svari}) as a block
type matrix
\begin{equation}
X = \pmatrix{A & B\cr C & D}~, \quad
\begin{array}{l}
A={-\bar{m}\over \Box_{+}}{D^{2}\bar{D}^{2}\over \Box}(g\Phi+M
Q^2)~, \; B =
{D^{2}\over\Box_{+}}\bar{g}\bar{\Phi}~, \\
D={-m\over \Box_{+}}{\bar{D}^{2}D^{2}\over
\Box}\bar{g}\bar{\Phi}~, \; C = {\bar{D}^{2}\over\Box_{+}}(g\Phi +
MQ^2 )~.
\end{array}
\end{equation}

The trace of the first power of the matrix doesn't contain the
field dependence. One can see that divergent terms $\sim M$ will
appear only in 2,
3 and 4-th orders. We will consider them one after another.\\
In the second order one gets $-\frac{1}{2}\cdot 2M
D^2\bar{D}^2(\bar{m}^2g\Phi\frac{1}{\Box \Box_+^2}
+\bar{g}\bar{\Phi}\frac{1}{\Box_+^2})Q^2$
\begin{equation}
{\rm Tr}(X^2)/2 \sim -\frac{1}{2(4\pi)^2} M (\bar{m}^2 g \Phi +
2\bar{g}\bar{\Phi}m \bar{m}) \ln (\frac{m\bar{m}}{L^2})~.
\end{equation}
The third order gives
$\frac{-\bar{m}}{\Box_+^3}D^2\bar{D}^2
\bar{g}\bar{\Phi}2 g\Phi MQ^2 + \frac{D^2\bar{D}^2}{\Box_+^3}(-m
\bar{g}^2\bar{\Phi}^2MQ^2 )$
\begin{equation}
{\rm Tr}(X^3)/3 \sim -\frac{1}{2(4\pi)^2}(\bar{m}
2\bar{g}g\Phi\bar{\Phi} + m \bar{g}^2 \bar{\Phi}^2 )M
\ln(\frac{m\bar{m}}{L^2})
\end{equation}
The fourth order gives
$-D^2\bar{D}^2\frac{\Box}{\Box_+^4}\bar{g}^2\bar{\Phi}^2 g\Phi M
Q^2$
\begin{equation}
{\rm Tr}(X^4)/4 =-\frac{1}{2(4\pi)^2} g \bar{g}^2 \Phi
\bar{\Phi}^2 M \ln(\frac{m\bar{m}}{L^2})
\end{equation}
Here $L$ is a regularization scale. Combining these terms together
one obtains the following expression
\begin{equation}\label{ldiv}
W_{\rm div}= -\frac{1}{2(4\pi)^2} \{\bar{m}^2 g \Phi
+2m\bar{g}\bar{G} + 2 g\bar{g}\Phi \bar{G}\}M
\ln(\frac{m\bar{m}}{L^2}),
\end{equation}
where $\bar{G}= -D^2 \Phi= Q^2 \Phi$ is the earlier introduced
combination, which was used in the equations of motion
(\ref{eqofmot}). The result (\ref{ldiv}) corresponds to the one
given in \cite{g} up to regularization scheme.

The second approach for calculation of divergencies implies use of
the expansion (\ref{intexp}). To get the divergencies it is enough to
consider in the sum of (\ref{expan}) only terms coming from $R_{0}$:
\begin{equation}
R_{0}^{\rm div} = \int d^4x d^2\theta\,
\frac{m}{2(4\pi)^2}(\bar{m}^2+2\bar{g}Q^2\Phi)(\frac{1}{\lambda}+
hQ^2\Phi)\ln(\frac{m\bar{\mu}}{L^2})~,
\end{equation}
and $R_{1}$:
\begin{equation}
R_{1}^{\rm div} = \int d^4x d^2\theta
\frac{g}{2(4\pi)^2}(\bar{m}^2+2\bar{g}Q^2\Phi)(\frac{1}{\lambda}+
h Q^2\Phi)\Phi \ln (\frac{m\bar{\mu}}{L^2})~.
\end{equation}
In the above expressions we used relations
$(\bar{m}+\bar{g}\bar{A})^2= \bar{m}^2 +2\bar{g} \bar{G}$ and
$\bar{G}=Q^2\Phi$.  The sum of $R_{0}$ and $R_{1}$ gives the
result which coincides with (\ref{ldiv}) up to finite terms and
divergent terms independent of $\Phi$. We point out also that the
expression $W_{div}$ (\ref{ldiv}) is completely consistent with
all earlier results on one-loop divergencies in the model under
consideration obtained by the other methods \cite{g}.

\subsection{Structure of finite contributions to chiral effective
potential}\label{structure} In this section we demonstrate using
(\ref{intexp}) the explicit calculations of several finite
contributions to effective potential with higher orders of
external grassmannian momenta.  According to the previous
discussion, these terms can be presented as elements of a ring
structure $R(\Phi, D^{2}\Phi, D\Phi D\Phi)$ (see also Ref.
\cite{ss}). From (\ref{sume}, \ref{sampl}, \ref{aint}) one can see
that the terms $R_{n}$ in the expansion (\ref{expan}) should have
the general form
\begin{equation}\label{gener}
R_{n} = \int d^4x d^2\theta\, \sum_{k=0}^{n} C_{k;n}(M, \bar\mu,
m)\Phi^{n-k}({M\over m}D^2\Phi)^k~, \quad \Phi = \Phi(y, \theta)
\end{equation}
here as before $M= hQ^2 \Phi + \frac{1}{\lambda}$, $\bar{\mu}
=\bar{m}+\bar{g}\bar{\Phi}$, and $C_{k;n}(M, \bar\mu, m)$ are some
functions. It can be also mentioned from (\ref{intexp}) that
$\Gamma_{(1)}$ is proportional to $M= hQ^2 \Phi + \frac{1}{\lambda}$
and at $h=0$ and $1/\lambda = 0$ we get $\Gamma_{(1)} = 0$, i.e. then
${\cal N}=1$ supersymmetry is recovered, the chiral effective potential
is absent in agreement with the nonrenormalization theorem.

We will show that calculations of the finite contributions $R_{n}$
to the chiral effective potential is closely related with so
called Mellin-Barnes representation of hypergeometrical functions
of several variables \cite{hyper}. The hypergeometrical functions
of grassmannian variables are the finite order polynomials which
can be always expressed via a star-product. It means that the
effective potential can be organized as a sum of elements of the
mentioned ring.

\subsubsection{$R_{2}$ contribution to the effective potential}
For the beginning one notes that all terms $R_{n}$ for $n>1$ in
(\ref{intexp}) are finite. We will consecutively consider several
first finite contributions to the chiral effective potential. In
this section we find an explicit expression for $R_{2}$ term from
the expansion (\ref{expan}) which defined by $E_{(2)}$
contribution in Eq. (\ref{intexp}). Since the expansion of
$E_{(2)}$ contains the momenta variables, after the inverse
Fourier transform the result will contain terms with derivatives,
i.e. $\pi_i\Phi_i\rightarrow D\Phi$. Of course the result contains
a constant contribution which is summed in (\ref{colwein}), but it
also contains the contributions with derivatives of the background
fields.

A method of $R_{n}$ terms calculation was described in the
beginning of section \ref{eval}.  It supposes using the $E_{(n)}$
expansion in the form (\ref{sampl}). However one can start
directly with the exponential form (\ref{expon}).

In the expansion (\ref{expan}) we consider $R_{2}$ contribution
containing one variable $l_{12}= \pi_{2}$. According to
(\ref{intexp}) and (\ref{sume}) it will be defined by $E_{(2)}$,
which is the function of one variable $l_{12}$, as it demonstrated
in (\ref{sampl}). If we consider the Mellin transform (see e.g.
\cite{hyper}) for functions dependent on one variable
\begin{equation}
f(\rho)=\int_0^{\infty} f(l^2_{i})(l^2_{i})^{\rho
-1}dl^2_{i}~,\quad f(l^2_{i})=\int_{C_{-}} \frac{d\rho}{2\pi i}
f(\rho)(l^2_{i})^{-\rho}~,
\end{equation}
we can note that $R_{(2)}$ is the Mellin
transform of function $E_{(2)}$, with $l_{i}=l_{12}$,
$f(\rho)=\frac{\Gamma(\rho)}{(-T M \alpha_1\alpha_2)^{\rho}}$.
The integration contour $C_{-}$ in the complex $\rho$-plane goes
from $-\infty$ and must separate "right" set of poles in the
integrand from "left" set of poles. Further we will understand all
contour integrals in this sense.

After integration over the proper-time $T$ and $\alpha_{1},
\alpha_{2}$ one obtains the expression:
\begin{equation}
R_{2}= \frac{g^2}{4(4\pi)^2}\cdot\frac{\bar{\mu}^2 M}{m}\int
d^4x\int d\pi_{1}d\pi_{2}\,\Phi(y, \pi_{1}) \Phi(y, \pi_{2})
\delta(\pi_1+\pi_2)I_{2}~,
\end{equation}
where
\begin{equation}
I_{2} = \left.\left(\frac{d}{ds}\right)\right|\int_{C}{d\rho\over
2\pi i} \,(-l_{12}^2\frac{M}{m})^
{-\rho}\frac{\Gamma(s+1-\rho)}{\Gamma(s)}
\frac{\Gamma(\rho)\Gamma^2(1-\rho)}{\Gamma(-2\rho +2 )}~.
\end{equation}
Here $l_{12}=\pi_{1}$ (see denotation (\ref{l})) and
$\Gamma(\rho)$ the Euler gamma function \cite{hyper}. Implementing
$\frac{d}{ds}|$ and using the known properties of $\Gamma(2 z)$
one obtains
$$ I_{2} =\int\frac{d\rho}{2\pi
i}(-l^2_{12}\frac{M}{m})^{-\rho}\frac{\sqrt{\pi}}{2}
\frac{\Gamma^2(1-\rho)\Gamma(\rho)}{\Gamma(\frac{3}{2}-\rho)}~,
$$
which is a representation for the hypergeometric
function\footnote{To see this, one should change $\rho \rightarrow
-\rho$ and $C_{-} \rightarrow C_+$.}
$_{2}F_{1}(1,1;\frac{3}{2};z)$ (see e.g. \cite{hyper}). Finally we
can rewrite this contribution as follows
\begin{equation}
\begin{array}{c}
R_{2}= \frac{g^2}{4(4\pi)^2}\cdot \int d^4x d^2\theta \frac{M}{m}
\bar{\mu}^2 \Phi(y, \theta) \,
_{2}F_{1}(1,1;\frac{3}{2};\frac{M}{4m}D^2)\Phi(y, \theta) =\\
= \frac{g^2}{4(4\pi)^2}\int d^4x d^2\theta\, \frac{M}{m}
\bar{\mu}^2 \left(\Phi^{2} + {M\over 6m}\Phi D^{2}\Phi\right)~,
\end{array}
\end{equation}
we remind that $\bar\mu =\bar{m}+\bar{g}\bar{\Phi}$, $M= h(Q^2
\Phi) + \frac{1}{\lambda}$ and use the explicit expression for
$_{2}F_{1}(1,1;\frac{3}{2};z)=\frac{\arcsin\sqrt{z}}{\sqrt{z(1-z)}}
= 1+\frac{1}{3!}\frac{M}{m}l^2$ where $z=l_{12}^2\frac{M}{m}$ in
the given case.

In bosonic sector this contribution has the simple form
\begin{equation}
R_{2}^{b}= \frac{g^2}{4(4\pi)^2}\int d^4x\, \frac{M}{m} \bar{m}^2
\left(2AF + {M\over 6m}F^2\right)~, \quad M= hF +
\frac{1}{\lambda}~.
\end{equation}

\subsubsection{$R_{3}$ contribution to the effective potential}
It is remarkable that all terms $R_{(n)}$ in the expansion (\ref{expan})
can be founded using the method demonstrated
in the previous section (i.e. using the Mellin transform).
For higher $R_{n}$ the only difference concerns only number of
variables.

In the present section we propose a general method for calculation
of $R_{n}$ contributions based on the mentioned Mellin-Barnes
representation for hypergeometric functions of several variables
\cite{hyper}. In order to illustrate the main idea, we consider
contribution to the effective potential from $R_{3}$, which
contains three variables. Substituting the expression $E_{3}$
given in (\ref{sampl}) to general expansion (\ref{intexp}) one
obtains
\begin{equation}\label{3leg1}
\begin{array}{rl}
R_{3} =& \frac{g^3 M}{6(4\pi)^2}\cdot\frac{\bar{\mu}^2}{m^2}\int
d^4x \int \prod d^2\pi\, \Phi(y, \pi_1)\Phi(y, \pi_2)\Phi(y,
\pi_3) \,\delta(\sum_{i}
\pi_{i})\,I_{3}~,\\
 I_{3} =& \frac{\sqrt{\pi}}{4}\int \frac{d^3\rho}{(2\pi
i)^3}(-l^2_{12}\frac{M}{4m})^{\rho_1}
(-l^2_{13}\frac{M}{4m})^{\rho_2}(-l^2_{23}\frac{M}{4m})^{\rho_3}\times\\
&\times \Gamma(-\rho_1)\Gamma(-\rho_2)\Gamma(-\rho_3)
\frac{\Gamma(\rho_1+\rho_2+1)\Gamma(\rho_1+\rho_3+1)\Gamma(\rho_2+\rho_3+1)}{\Gamma(\rho_1+\rho_2+\rho_3+\frac{3}{2})}~,
\end{array}
\end{equation}
where $l_{ik}$ are defined by (\ref{l}):
$l_{12}^2=\pi_2^2=-\frac{1}{2} \pi_2 (\pi_1 + \pi_3)$, $l_{13}^2 =
\pi_1^2 =-\frac{1}{2}\pi_1 (\pi_2 + \pi_3)$,
$l_{23}^2=\pi_3^2=-\frac{1}{2} \pi_3 (\pi_1 + \pi_2)$, and $I_{3}$
is the Mellin-Barnes representation for hypergeometric functions
of several variables (three variables in the case under
consideration) \cite{hyper}.  It should be noted that variables
$l_{12},\,l_{23},\, l_{13}$ appear in the expression (\ref{3leg1})
symmetrically. Using the definition for the hypergeometric
function ${_{2}F}_{1}$ of two variables \cite{hyper} one can write
\begin{equation}\label{i3x} I_{3}=
\sum_{n_1,n_2=0}^{\infty}\frac{\sqrt{\pi}}{4}(l_{12}^2
\frac{M}{4m})^{n_1}(l_{13}^2\frac{M}{4m})^{n_2}
\frac{\Gamma(n_1+n_2+1)}{\Gamma(n_1+n_2+\frac{3}{2})}\times
\end{equation}
$$
\times\,{_{2}F}_{1}(n_1+1,n_2+1;n_1+n_2+\frac{3}{2};l_{23}^2\frac{M}{4m})~.
$$

Properties of hypergeometric functions allows us to rewrite the
last expression as follows
\begin{equation}\label{i3}
I_{3}= \frac{\sqrt{\pi}}{4}\sum_{n=0}^{\infty}(u)^n
\frac{\Gamma(n+1)}{\Gamma(n+\frac{3}{2})}F_1(1,n+1,n+1;n+\frac{3}{2};s,t)~,
\end{equation}
where
$$
u=l_{12}^2\frac{M}{4m}~,\quad s=l_{23}^2\frac{M}{4m}~,\quad
t=l_{13}^2\frac{M}{4m}~.
$$
The definitions for $l_{kn}$ are given in (\ref{l}).
Further we use $F_1$-type function or the
Appel's hypergeometrical function from, so called, the Horn list of
functions \cite{hyper}.

Using the property $(l_{i_{1}i_{2}}^{2})^{k}= 0,\, k>1$, one can
rewrite Eq. (\ref{i3}) as a finite order polynomial. It leads for
the $3$-legs contribution to
\begin{equation}
\begin{array}{rl} R_{3} =
&\frac{-g^3}{6(4\pi)^2}\,(\frac{M}{m^2})\bar{\mu}^2 \int d^4x\int \prod
d^{2}\pi \, \Phi(y, \pi_{1})\Phi(y, \pi_{2}) \Phi(y, \pi_{3})
\delta(\sum_{i} \pi_{i})\times\\ &\times \left(\frac{1}{2} +
\frac{1}{3\cdot4}(s+t+u)+\frac{1}{3\cdot4\cdot5}(st+su+tu)
+\frac{1}{5\cdot6\cdot7}stu\right)~.
\end{array}
\end{equation}
Taking into account the definitions for $u, t, s$, the definitions
(\ref{pi}), the expansion (\ref{sume}) and discussion concerning
the inverse Fourier transform given after Eq. (\ref{aint}), we
transform the expression for $R_{3}$ in the coordinate
representation
\begin{equation}
\begin{array}{c} R_{3} = \frac{g^3}{6(4\pi)^2}\,\int d^4x
d^2{\theta}\,(\frac{M}{m^2})\bar{\mu}^2\times\\
\times \left(\frac{1}{2}\Phi^{3} +
\frac{1}{4}\frac{M}{4m}\Phi^{2}D^{2}\Phi+\frac{1}{4\cdot5}\left(\frac{M}{4m}\right)^2\Phi
D^{2}\Phi D^{2}\Phi+
\frac{1}{5\cdot6\cdot7}\left(\frac{M}{4m}\right)^3(D^{2}\Phi)^{3}\right)~,
\end{array}
\end{equation}
where as before $\bar\mu =\bar{m}+\bar{g}\bar{\Phi}$, $M= hQ^2 \Phi +
\frac{1}{\lambda}$.

In the bosonic sector this result looks like
\begin{equation}
R_{3}^{b} = \frac{g^3}{6(4\pi)^2}\int d^4x\,\frac{M\bar{m}^2}{m^2}
 \left(\frac{3}{2}A^{2}F +
\frac{1}{2}\frac{M}{4m}AF^{2} + \frac{1}{4\cdot5}\left(\frac{M}{4m}\right)^2F^{3}\right)~,\;  M= hF + \frac{1}{\lambda}~.
\end{equation}

It is easy to understand that exploiting the Mellin transform
for higher $R_{n}$ contributions, we will obtain the expressions
containing the hypergeometric function of several variables.
Such functions for $n>2$ are called generalized Lauricella hypergeometric
functions \cite{hyper, 1loop}. This is the expected result. The general
expressions for all one-loop massive Feynman diagrams have been
studied in Refs. \cite{1loop} using the Mellin-Barnes
representation for hypergeometric functions of several variables
by contour integrals and it was shown that all of them can be
expressed via the Lauricella hypergeometric functions and the
described procedure, in principle, can be applied for calculations
of any higher $R_{n}$ contributions to chiral effective potential.

\subsection{Constant field contribution to the chiral
effective potential}\label{constant} In the previous sections we
considered two examples which demonstrated that, in principal, all
$R_{n}$ can be calculated and the results can be expressed via
hypergeometric functions of several variables. Now we construct an
approximation taking into account some terms from all $R_{n}$
Since every $R_{n}$ is a finite order polynomial in powers of
$D^{2}\Phi$ (see the discussions in the beginning of section
(\ref{structure}) we can try, as a first approximation, to sum up
all terms without these derivatives and obtain an approximate
expression for the chiral effective potential containing the
contributions from all $R_{n}$.

The expansion (\ref{sume})\footnote{See also examples
(\ref{sampl}).} always begins from constant, i.e. $E_{(n)} = 1
+\cdots $. Let's choose from all $E_{(n)}$ only the terms without
variables $l_{ij}$ and calculate their contribution to
(\ref{intexp}). It will correspond to the situation then we take
into account only $k=0$ term in the representation (\ref{gener})
for all $R_{n}$, i.e. we sum contributions from all $R_{n}$ which
have no grassmannian derivatives.

One can see that the sum of contributions to the chiral effective
potential (\ref{intexp}) containing no Grassmann derivatives is
written in the following form
\begin{equation} \Gamma_{(1)}^{(0)}=
\frac{1}{2(4\pi)^2} \int d^4x d^2\theta\, m\bar{\mu}^2 M
\sum_{n=2}^{\infty}\left(-\frac{g\Phi(y, \theta)}{m}\right)^n
\frac{1}{n(n-1)}~,
\end{equation}
where multiplier $\frac{1}{n(n-1)}$ appears as the result of the
proper time integration in (\ref{intexp}).

 The sum can be calculated exactly, that gives
\begin{equation}\label{colwein}
\Gamma_{(1)}^{(0)}= \frac{1}{2(4\pi)^2} \int d^4x d^2\theta\,
m(\bar{m}+\bar{g}\bar{\Phi})^2 (hQ^2 \Phi +
\frac{1}{\lambda})\left(-\frac{g\Phi}{m}+(1+\frac{g\Phi}{m})
\ln(1+\frac{g\Phi}{m})\right)~.
\end{equation}
The expression (\ref{colwein}) is the chiral effective potential
in approximation when all terms containing the $D^2{\Phi}$ can be
neglected, however all terms without these derivatives are exactly
summed up. The corrections to this approximation obligatory
contain the terms with Grassmann derivatives of the background
field.

Now we study a structure of the effective potential
(\ref{colwein}) in the bosonic component sector. We have
\begin{equation}
\Gamma_{(1)}^{(0)}|_b= \frac{m\bar{m}^2}{2(4\pi)^2}\int d^4x\, (hF +
\frac{1}{\lambda}) {g\over m}F\ln(1+\frac{g}{m}A)~.
\end{equation}
This expression gives a correction to the classical potential
(\ref{bpot})
\begin{equation}\label{delta}
\Delta V =
\frac{m\bar{m}^2}{2(4\pi)^2} (hF + \frac{1}{\lambda}) {g\over
m}F\ln(1+\frac{g}{m}A)~.
\end{equation}
Eliminating the auxiliary field from the classical equations of
motions (\ref{mot}) one finds the full one-loop effective
potential in this approximation
\begin{equation}\label{final}
V^{(1)}= V + \Delta V =
\bar{G}(G
+\frac{1}{\lambda}\bar{G}+\frac{h}{6}\bar{G}^2)+\frac{g\bar{m}^2}{2(4\pi)^2}
(-2h\bar{G} + \frac{1}{\lambda})
\ln(1+\frac{2g}{m^2}G)~,
\end{equation}
where $G=m A+ \frac{g}{2}A^2, \; \bar{G}=\bar{m} \bar{A}+
\frac{\bar{g}}{2}\bar{A}^2$. The exact one-loop effective
potential includes, besides (\ref{colwein}), the powers of
$D^2{\Phi}$. It means, the exact component form of the effective
potential will contain, besides (\ref{delta}) the powers of $F$.
Therefore, after eliminating the $F, \bar{F}$-terms from classical
equations of motion, one gets the powers of $G, \bar{G}$ as
correction terms in (\ref{final}).

\section{Summary}
We have developed the general approach for finding the one-loop
effective potential in ${\cal N}=\frac{1}{2}$ noncommutative
Wess-Zumino model. Using the symbol-operator techniques we
obtained a general expression for the effective potential in terms
of a superfield heat kernel. This heat kernel was exactly
calculated for the specific background superfields determining the
functional dependence of  the chiral effective potential. As a
result we found the exact form of the effective potential
including the complete dependence on $\Phi$ and $D^2{\Phi}$ in
term of a single proper time integral. This effective potential
contains all previously obtained results for the theory under
consideration as partial cases.

To clarify the structure of the effective potential in more
details we have constructed a systematic procedure of expansion of
the effective potential in a power series over $\Phi$ and
$D^2{\Phi}$. Each term of this expansion can be calculated in an
explicit form. We demonstrated the calculations of divergent
contributions to the one-loop effective potential as well as a few
first finite contributions. All finite contributions to the
effective potential are expressed in terms of hypergeometrical
functions of several variables.

The expansion of the effective potential has an enough simple
structure and allows to organize resummation of the above series
and to get a series in derivatives $D^2{\Phi}$ with the
coefficients depending on $\Phi$. We have demonstrated how to
obtain the first term in this new expansion containing no
derivatives but including all powers of $\Phi$.

To conclude, we carry out the complete analysis of the one-loop
effective potential in ${\cal N}=\frac{1}{2}$ WZ model. The final
exact solution for this effective potential is constructed and
various its approximate forms are found.

\section*{Acknowledgements}
The work was supported in part by INTAS grant, INTAS-00-00254 and
RFBR grant, project No 03-02-16193.  I.L.B is grateful to RFBR
grant, project No 02-02-04002, to DFG grant, project No 436 RUS
113/669, to LRSS grant, project No 1252.2003.2 and to INTAS grant,
INTAS-03-51-6346 for partial support.  The work of N.G.P and A.T.B was
supported in part by RFBR grant, project No 02-02-17884.

\bibliographystyle{JHEP}

\end{document}